\documentclass{ws-ijmpe}

\newcommand{\rhosat}{\rho_{0}}                         
\newcommand{\fm}{\,\text{fm}}                          
\newcommand{\MeV}{\,\text{MeV}}                        
\newcommand{\pF}{p_{\text{F}}}	                       
\newcommand{\sclgth}{a}                                
\newcommand{\reff}{r_{\text{ef\mbox{}f}}}              
\newcommand{\etal}{\textit{et al.,\;}}
\newcommand{\Tc}{T_{c}}                                
\newcommand{\bvec}[1]{\textbf{#1}}                     
\newcommand{\bvecp}[1]{\textbf{#1}^{\prime}}           
\newcommand{\avg}[1]{\langle #1 \rangle}               
\newcommand{\createOpR}[2]{\hat{\psi}^{\dagger}_{#1}(#2)} 
\newcommand{\anihilOpR}[2]{\hat{\psi}_{#1}(#2)}        
\newcommand{\gTwo}[1]{g_{2}(#1)}                       
\newcommand{\gOne}[1]{g_{1}(#1)}                       
\newcommand{\IntR}[1]{\int d^{3}#1\;}                  
\newcommand{\DoubleIntR}[2]{\int d^{3}#1 d^{3}#2\;}    
\newcommand{\Ns}{N_{s}}                                
\newcommand{\eF}{\varepsilon_{\text{F}}}               
\newcommand{\NNtau}{n_{\tau}}                          
\newcommand{\TGreen}{{\cal G}}                         
\newcommand{\DeltaMin}{\Delta_{\textrm{min}}}          

\begin{document}

\markboth{G. Wlaz{\l}owski, P. Magierski}{Superfluid properties of dilute neutron matter}

\catchline{}{}{}{}{}

\title{SUPERFLUID PROPERTIES OF DILUTE NEUTRON MATTER}
\author{\footnotesize GABRIEL WLAZ{\L}OWSKI and PIOTR MAGIERSKI}

\address{Faculty of Physics, Warsaw University of Technology, ul. Koszykowa 75\\
00-662 Warsaw,
Poland\\
e-mail: gabrielw@if.pw.edu.pl, magiersk@if.pw.edu.pl}

\maketitle

\begin{history}
\received{(received date)}
\revised{(revised date)}
\end{history}

\begin{abstract}
We report results of fully non-perturbative calculations, based on Auxiliary Field Quantum Monte Carlo (AFQMC) approach, for the dilute neutron matter at the density $\rho=0.003\fm^{-3}$. 
Fundamental quantities which characterize the superfluid state: the single particle 
energy gap $\Delta(T)$, and the critical temperature $\Tc$ have been determined. 
The large value of $\Delta(0)/\Tc\approx 3.2$ indicates that the system is not a BCS-type superfluid 
at low temperatures.
\end{abstract}

\section{Introduction}
Although a homogeneous neutron matter is one of the simplest nuclear systems, its importance cannot
be overemphasized since it constitutes the main component of neutron stars. 
The density of the neutron matter forming neutron stars ranges from subnuclear densities (in the inner crust) up to extremely high values of $\rho\sim 10\,\rhosat$ (expected in the center of the star), 
where $\rhosat=0.16\fm^{-3}$ is the saturation density\cite{NS}.

In the regime of sufficiently low densities one can perform very precise calculations 
since the neutron-neutron interaction is completely dominated by the scattering in $^{1}S_{0}$ channel,
which is determined by
two parameters only: the scattering length $\sclgth$ and the effective range $\reff$. Indeed, at the densities 
$\rho \lesssim 0.03\,\rhosat$ ($\pF\lesssim 0.6\fm^{-1}$)
the influence of other channels as well as of three-body forces is marginal and can be neglected\cite{body3}. The values of the scattering length and the effective range are well known from the low energy scattering experiments and read: $\sclgth=-18.8\pm0.3\fm$, $\reff=2.75\pm0.11\fm$\cite{ScattExp}. 
At the densities where $| \pF \sclgth | \gtrsim 1$ the dilute neutron matter is an example of a strongly correlated system. Moreover, the influence of the effective range cannot be ignored except for very low densities
where $\pF\reff << 1$\cite{SchwenkPethick}. It implies that only non-perturbative approaches are able to gain a reliable insight into physics of this system. The large class of such methods, which are known under the general name of Quantum Monte Carlo (QMC), has been applied for the dilute neutron matter\cite{FantoniEtAl,GandolfiEtAl,GezerlisCarlson}. 
In spite of a considerable
theoretical effort some of its static properties are still vaguely known. 
The open questions mostly concern the superfluid state, which is generated at sufficiently 
low temperatures. For example the predicted value of the zero temperature energy gap 
differs significantly for various approaches, revealing
a strong dependence on the details of the applied method (see\cite{GandolfiEtAl} and references therein). 
The same concerns the critical temperature of the superfluid-normal phase transition, where 
the well known BCS formula $\Delta(0)/\Tc\approx 1.76$ is believed to be valid.

Here we present the selected results concerning the superfluid state, obtained within the AFQMC approach 
at non-zero temperatures\cite{bdm,unitary_review} for the dilute neutron matter of density $\rho = 0.02\,\rhosat$ ($\pF= 0.45\fm^{-1}$). This study can be regarded as a continuation of the
investigations presented in the papers \cite{WlazlowskiMagierski}, where the details of the AFQMC 
algorithm and the procedure of adjusting the interaction, have been discussed. 
In the present study we focused on two basic quantities which characterize a superfluid state: 
the critical temperature and the pairing gap.

\section{Critical temperature}
In the case of superfluidity the convenient order parameter is specified by 
the long-distance behavior of the two-body density matrix 
$
 \rho_{2}(\bvecp{r}_{1},\bvecp{r}_{2},\bvec{r}_{1},\bvec{r}_{2})=\avg{
	\createOpR{\uparrow}{\bvecp{r}_{1}}\createOpR{\downarrow}{\bvecp{r}_{2}}
  \anihilOpR{\downarrow}{\bvec{r}_{2}}\anihilOpR{\uparrow}{\bvec{r}_{1}}
},
$
where the field operators $\createOpR{\lambda}{\bvec{r}}$ and $\anihilOpR{\lambda}{\bvec{r}}$ obey the fermionic anticommutation relations. Such a long-distance behavior is directly related to the condensate fraction $\alpha$, which measure the fraction of Cooper pairs forming a condensate\cite{Yang}. 
This quantity can be efficiently computed within AFQMC, provided the information concerning 
the one-body density matrix 
$
 \rho_{1}(\bvec{r}_{1},\bvec{r}_{2})=\avg{
	\createOpR{\uparrow}{\bvec{r}_{2}}\anihilOpR{\downarrow}{\bvec{r}_{1}}
}
$
is also included. The condensate fraction is given by:\cite{Astrakharchik}
\begin{equation}
  \alpha=\lim_{r\rightarrow\infty} h(r),\qquad h(r)=\dfrac{N}{2}\gTwo{r}-g_{1}^{2}(r),\label{eqn:alpha}
\end{equation} 
where $N$ stands for the number of particles and the correlation functions $g_{1,2}(r)$ are defined as:
\begin{eqnarray}
 \gOne{\bvec{r}}&=&\dfrac{2}{N}\IntR{\bvec{r}_{1}}\rho_{1}(\bvec{r}_{1}+\bvec{r},\bvec{r}_{1}),\\
 \gTwo{\bvec{r}}&=&\left(\dfrac{2}{N} \right)^{2}\DoubleIntR{\bvec{r}_{1}}{\bvec{r}_{2}}
	\rho_{2}(\bvec{r}_{1}+\bvec{r},\bvec{r}_{2}+\bvec{r},\bvec{r}_{1},\bvec{r}_{2}).
\end{eqnarray}

\begin{figure}[t]
 \centering
 \includegraphics[scale=0.49]{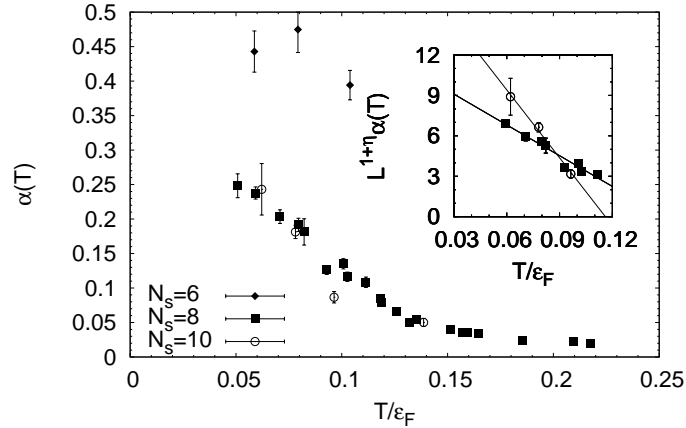}
 \caption{
The condensate fraction $\alpha$ as a function of temperature (in units of Fermi energy $\eF$)
for three lattice sizes $\Ns$. The inset presents the rescaled condensate fraction obtained 
for lattices $\Ns=8$ (squares) and $\Ns=10$ (circles). The intersection
of the fitted lines determines the critical temperature of the superfluid-normal phase transition, $\Tc\approx0.09\,\eF$.
}
 \label{fig:nm_alpha}
\end{figure}
The figure \ref{fig:nm_alpha} presents the condensate fraction versus the temperature for three different lattice sizes $\Ns$. It is clearly visible that the results for $\Ns=6$ significantly deviate from 
those obtained for larger lattices. It is
due to the fact that for small lattices the limiting value of $h(r)$
is rather poorly determined. Indeed, ''the infinity'' in this case
is equal to the half of the box size and is only a few times larger 
than the interaction range. Consequently, it will generate a large
systematic error when the lattice is too small. The presented results
indicate that this is exactly the case for $\Ns=6$ and therefore
these data were not included in the process of determination of the critical temperature. 

In the thermodynamic limit 
the condensate fraction has to vanish at the critical temperature $\Tc$. However, in the
case of calculations in the box the finite size effects come into play and smooth out 
all singularities typical for the phase transition.
Still, one can estimate the critical temperature using the method based on the finite size scaling.
Similar technique has been used to determine the critical temperature for cold atomic gases
(see\cite{bdm,BurovskiEtAl} for details). 
The volume-dependent estimation of the critical temperature $\Tc^{(ij)}$ is obtained by finding 
the intersection 
of the rescaled condensate fraction $L^{1+\eta}\alpha$ for two different lattice sizes $N_{i,j}$, where  $\eta\simeq 0.038$ is the universal critical exponent and $L^{3}$ is the volume of the box.
As $N_{i,j}\rightarrow\infty$ the thermodynamic limit is recovered and the series 
$\Tc^{(ij)}$ converges to the critical temperature. To extract 
$\Tc$ we have used two largest of the available lattices, namely $N_{i,j}=8,10$. 
The filling factor in both cases reads: $\nu=N/2\Ns^{3}\approx5\%$.
According to our experience, the above methodology is sufficient to estimate the critical temperature 
with an uncertainty smaller than $20\%$ (in fact this procedure applied to the unitary gas provides the value of $\Tc$ 
with the relative error smaller than $10\%$). Finally, the estimated value of the critical temperature reads 
$\Tc\approx 0.09\,\eF$ ($0.39\MeV$), where $\eF$ is the Fermi energy (see the inset of Fig. \ref{fig:nm_alpha}).

\section{Pairing gap}
In order to determine the gap in the single particle spectrum,  
we have computed the spectral weight function $A(\bvec{p},\omega)$.
This quantity has been extracted from the imaginary time propagator ${\cal G}(\bvec{p},\tau)$ 
through the analytic continuation\cite{fw}.
The procedure is equivalent to solving the integral equation:
\begin{equation}
\TGreen(\bvec{p},\tau)=-\frac{1}{2\pi}\int_{-\infty}^{+\infty}
d\omega A(\bvec{p},\omega)\frac{\exp(-\omega\tau)}{1+\exp(-\omega\beta)},
\label{eqn:Ap}
\end{equation} 
where $\beta$ denotes the inverse of the temperature and 
$\TGreen(\bvec{p},\tau)$ is determined from the Monte Carlo calculations 
for the discrete set of values $\tau_{0}=0,\tau_{1},\ldots,\tau_{\NNtau}=\beta$.
Numerically, however, the above integral equation represents
an ill-posed problem i.e. there is an infinite class of solutions for $A(\bvec{p},\omega)$ which 
satisfy Eq. (\ref{eqn:Ap}) within uncertainties generated by Monte Carlo method. 
Therefore we have used two methods which were in particular designed to deal with such tasks:
the truncated SVD method and the maximum entropy method\cite{svd1,jaynes,MagierskiEtAl}. 
\begin{figure}[t]
 \centering
 \includegraphics[scale=0.30]{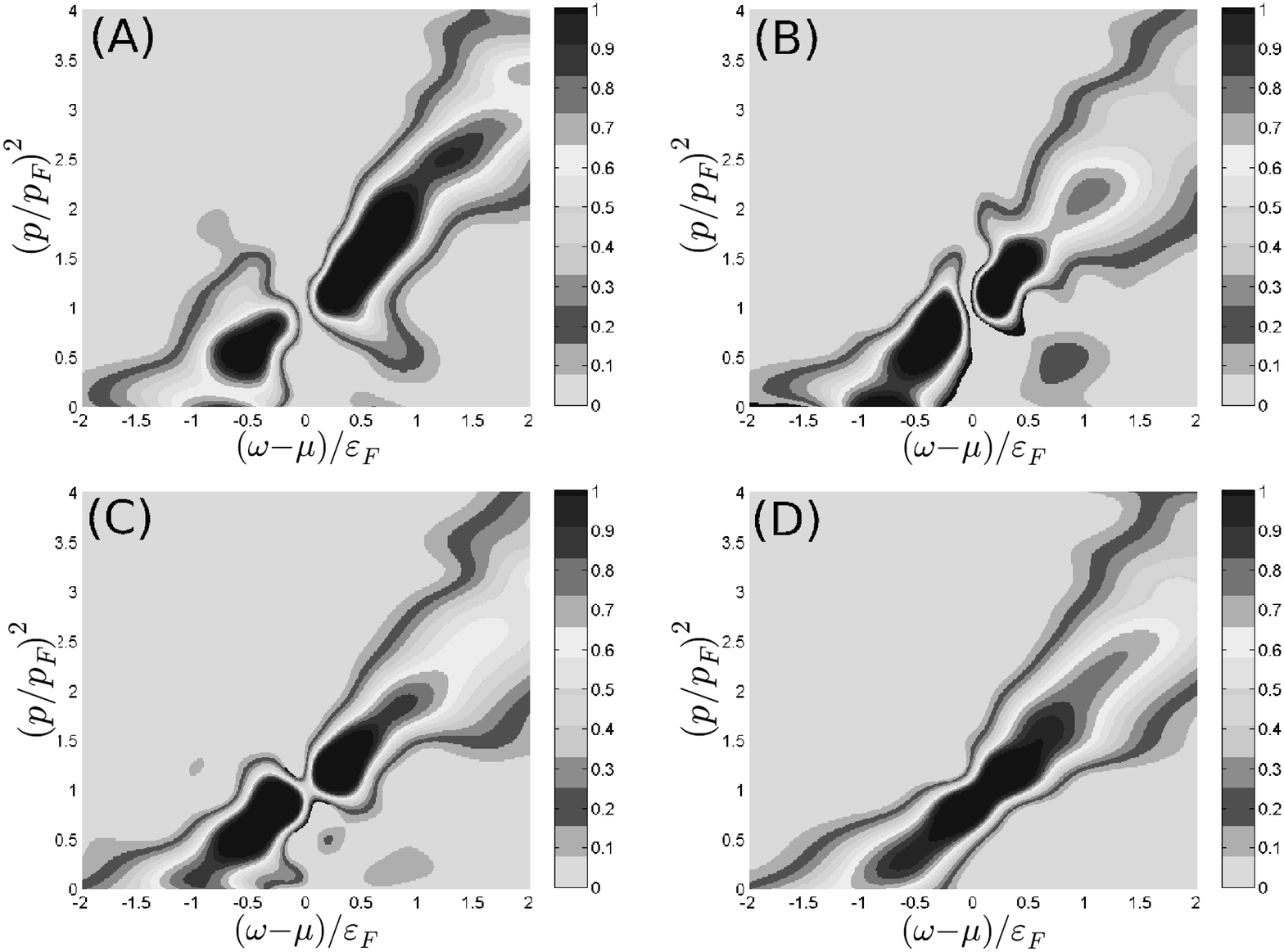}
 \caption{
The spectral weight function $A(\bvec{p},\omega)$ for the dilute neutron matter at selected temperatures: A)~$T\simeq 0.06\,\eF<\Tc$, B)~$T\simeq 0.08\,\eF\lessapprox\Tc$, C)~$T\simeq 0.10\,\eF\gtrapprox\Tc$, D)~$T\simeq 0.12\,\eF>\Tc$. 
The nonzero energy gap is present up to the critical temperature $\Tc\approx0.09\,\eF$.
}
 \label{fig:nm_spectral}
\end{figure}

The spectral functions for selected temperatures (below and above $\Tc$) are presented in Fig. \ref{fig:nm_spectral}. Our results admit the gapped solutions up to the critical temperature and above $\Tc$ the gap vanishes. 
One has to remember, however, that there is a finite resolution 
of both methods concerning the value of the gap. Namely, the gap cannot be resolved if its value decreases below $\DeltaMin\approx 0.2\,\eF$.

The value of the energy gap $\Delta$ extracted from the spectral function at the temperature $T\simeq 0.06\,\eF$ 
is expected to provide a good approximation of its value at zero temperature. 
It reads: $\Delta(0)/\eF=0.29^{+0.02}_{-0.04}$. 
\begin{figure}[t]
 \centering
 \includegraphics[scale=0.30]{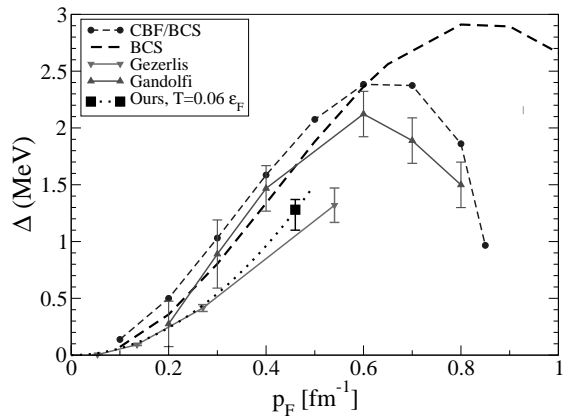}
 \caption{
Superfluid paring gap in various Monte Carlo calculations together
with results of pure BCS approach (dashed line) -  see text for details.
The dotted line corresponds to $\Delta(\pF)=0.29\,\pF^{2}/2m$.
}
 \label{fig:gap_comparison}
\end{figure}
In Fig.~\ref{fig:gap_comparison} our results are compared  
to those obtained by other authors within {\it ab~initio} calculations for zero temperature. 
We have found our results 
in agreement with the recent Green Function Monte Carlo calculations of Gezerlis and Carlson\cite{GezerlisCarlson}. The calculations including other scattering channels together with the three body forces
were performed by Gandolfi \etal\cite{GandolfiEtAl} and predict larger paring gap. 
However, it is difficult to establish the source of this discrepancy since their method 
(Auxiliary Field Diffusion Monte Carlo) has been constrained in order to avoid the fermionic sign problem 
and therefore can be regarded as a variational approach.

An interesting result can be noticed by 
considering the ratio of the energy gap at zero temperature to the critical temperature. 
Namely, the ratio $\Delta/\Tc\approx 3.2$ exceeds almost twice the well-known BCS value $1.76$. 
A similar situation has been encountered for the unitary gas, where the existence of the exotic ``pseudogap'' phase 
above $\Tc$ was recently reported\cite{MagierskiEtAl}. 
It indicates that the dilute neutron matter at this density is not a BCS-type superfluid.

\section{Conclusions}
We have performed the fully non-perturbative calculations for the dilute neutron matter of density $\rho\cong 0.02\rho_{0}$ at finite temperatures. We focused on the basic quantities which characterize the superfluid state and extracted the
value of the critical temperature and the pairing gap. Our results are free from uncontrolled
approximations and are essentially exact with only uncertainties related to statistical errors and finite size effects.
The large value of $\Delta/\Tc\approx 3.2$ suggests that the dilute neutron gas cannot be described by the BCS theory and thus is not a BCS superfluid at low temperatures.

\section*{Acknowledgments}
Support from the Polish Ministry of Science under contracts No. N N202 328234, N N202 128439 
and by the UNEDF SciDAC Collaboration under DOE grant DE-FC02-07ER41457
is gratefully acknowledged. Calculations reported here have been performed
at the Interdisciplinary Centre for Mathematical and Computational Modelling (ICM) at Warsaw University.

\end{document}